\documentclass[conference]{IEEEtran}
\IEEEoverridecommandlockouts

\usepackage{cite}
\usepackage{amsmath,amssymb,amsfonts}
\usepackage{algorithmic}
\usepackage{graphicx}
\usepackage{textcomp}
\usepackage{subcaption}
\usepackage{url}
\usepackage{xcolor}
\def\BibTeX{{\rm B\kern-.05em{\sc i\kern-.025em b}\kern-.08em
    T\kern-.1667em\lower.7ex\hbox{E}\kern-.125emX}}
\begin{document}

\title{Aligner-Guided Training Paradigm: Advancing Text-to-Speech Models with Aligner Guided Duration\\
}

\author{\IEEEauthorblockN{1\textsuperscript{st} Haowei Lou}
\IEEEauthorblockA{
\textit{UNSW Sydney}\\
Kensington, Australia \\
0009-0009-1359-872X}
\and
\IEEEauthorblockN{2\textsuperscript{nd} Helen Paik}
\IEEEauthorblockA{
\textit{UNSW Sydney}\\
Kensington, Australia \\
0000-0003-4425-7388}
\and
\IEEEauthorblockN{3\textsuperscript{rd} Wen Hu}
\IEEEauthorblockA{
\textit{UNSW Sydney}\\
Kensington, Australia \\
0000-0002-4076-1811}
\and
\IEEEauthorblockN{4\textsuperscript{th} Lina Yao}
\IEEEauthorblockA{
\textit{UNSW Sydney}\\
Kensington, Australia \\
0000-0002-4149-839X}
}

\maketitle

\begin{abstract}
Recent advancements in text-to-speech~(TTS) systems, such as FastSpeech and StyleSpeech, have significantly improved speech generation quality. However, these models often rely on duration generated by external tools like the Montreal Forced Aligner, which can be time-consuming and lack flexibility. The importance of accurate duration is often underestimated, despite their crucial role in achieving natural prosody and intelligibility. To address these limitations, we propose a novel Aligner-Guided Training Paradigm that prioritizes accurate duration labelling by training an aligner before the TTS model. This approach reduces dependence on external tools and enhances alignment accuracy. We further explore the impact of different acoustic features, including Mel-Spectrograms, MFCCs, and latent features, on TTS model performance. Our experimental results show that aligner-guided duration labelling can achieve up to a 16\% improvement in word error rate and significantly enhance phoneme and tone alignment. These findings highlight the effectiveness of our approach in optimizing TTS systems for more natural and intelligible speech generation.
\end{abstract}


\section{Introduction}
Text-to-speech~(TTS) systems have seen significant advancements with the development of state-of-the-art models such as FastSpeech~\cite{ren2019fastspeech,ren2020fastspeech} and StyleSpeech~\cite{lou2024stylespeechparameterefficientfinetuning}. These models feature efficient architectures that enable high-quality and natural-sounding speech generation. FastSpeech uses a non-autoregressive approach to generate speech quickly, while StyleSpeech incorporates style information to produce speech with diverse prosodic variations. However, both models share a common practice in their training paradigm: they rely on duration label to align phonemes with their corresponding speech segments. These durations are typically treated as ground truth and are generated using external tools.

The Montreal Forced Aligner (MFA)\cite{mcauliffe2017montreal} is a widely used tool for phoneme alignment, designed to align phonemes with their corresponding segments in speech using the Kaldi-ASR Toolkit~\cite{Povey_ASRU2011}. It automates the alignment process through a model based on Hidden Markov Models (HMMs)~\cite{rabiner1986introduction}, which efficiently processes phonetic transcriptions and audio recordings to generate time-aligned phoneme boundaries. These boundaries provide the duration labels that serve as the ground truth for training TTS models, ensuring that phonemes are accurately synchronized with the corresponding audio. The use of MFA has become a standard practice in the field, as it offers a convenient and reliable method to produce essential alignment data for TTS training.
\begin{figure}[t]
    \centering
    \includegraphics[width=0.6\linewidth]{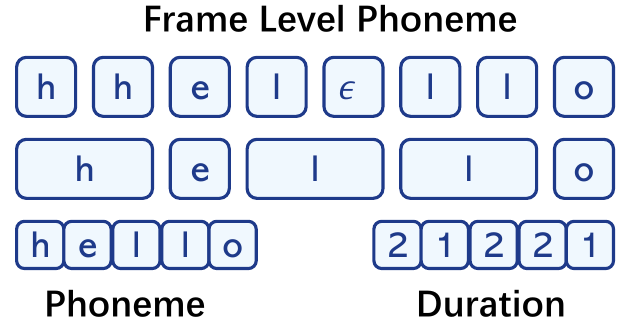}
    \caption{Phoneme Duration Alignment}
    \label{fig:ada}
\end{figure}
The MFA approach has facilitated the training of TTS models.  However, it presents two significant limitations to the current TTS training paradigm. First, the importance of accurate duration is often underestimated, with much of the focus on improving model architecture or other components. Precise duration alignment is crucial for achieving natural prosody and intelligibility; inaccuracies in duration labelling can negatively impact the quality of the generated speech. Second, the use of MFA, which relies on HMM-based approaches, involves long training times and lacks flexibility. This makes it less adaptable for rapidly evolving TTS development environments.

To address these limitations, this work proposes a novel approach that emphasizes the importance of duration in the training process. We introduce an Aligner-Guided Training Paradigm, where an aligner is first trained to generate accurate duration. These labels are then used to guide the training of the TTS model, ensuring better alignment and more natural speech generation. Our experimental results demonstrate that this approach significantly improves the performance of TTS models, achieving higher accuracy in phoneme and tone alignment and producing more natural-sounding speech. By leveraging aligner-guided duration labeling, our method provides a more efficient and flexible solution for training state-of-the-art TTS models.
The main contributions of this work are:

\begin{enumerate}
    \item We validate that duration plays a crucial role in TTS systems. Our results show that TTS models trained with accurate duration can achieve more than a 15\% improvement in speech accuracy.
    \item We propose a novel TTS training paradigm, called the \textbf{Aligner-Guided Training Paradigm}. It reduces reliance on external tools such as MFA and provides more reliable and accurate duration information.
    \item We explore the impact of different acoustic features, including Mel-Spectrograms, MFCCs, and latent features, on the performance of TTS models. Our empirical results show that Mel-Spectrograms significantly enhance TTS training and offer the best performance among the features studied. 
\end{enumerate}

\section{Method}
In this section, we present the \textbf{Aligner-Guided Training Paradigm} in detail. Given a speech signal $A$, a phoneme sequence $X$, and a target representation $Y$ for the TTS model. We first encode the speech signal into an acoustic feature~$H$, such as a Mel-Spectrogram~(MelSpec), Mel-frequency cepstral coefficients~(MFCC), or a latent feature. Next, we train an Automatic Speech Recognition~(ASR) model to recognise $X$ based on $H$ and employ an aligner module to calculate the duration $L$ for each phoneme in $X$. Finally, the TTS model is trained using the phoneme sequence $X$ and its corresponding duration $L$ to reconstruct the target output $Y$. Figure~\ref{fig:agt} presents an overview of the proposed Aligner-Guided Training Paradigm.

\subsection{Automatic Speech Recognition}
Assume we have a set of phonemes with a totally $P$ distinct phoneme, and a speech signal $A$. We first encode $A$ into an acoustic feature $H \in \mathbb{R}^{N \times T}$, where $T$ is the number of frames in the acoustic feature and $N$ is the frame dimension. The Automatic Speech Recognition~(ASR) model takes $H$ as input and generates a likelihood matrix, $C = \mathbf{ASR}(H) \in \mathbb{R}^{P \times T}$, where each element $C_{i,j}$ represents the likelihood of the $j$-th frame being associated with the $i$-th phoneme. 

\begin{figure*}[t]
    \centering
    \begin{subfigure}[t]{0.31\textwidth}
        \centering
        \includegraphics[width=0.5\linewidth]{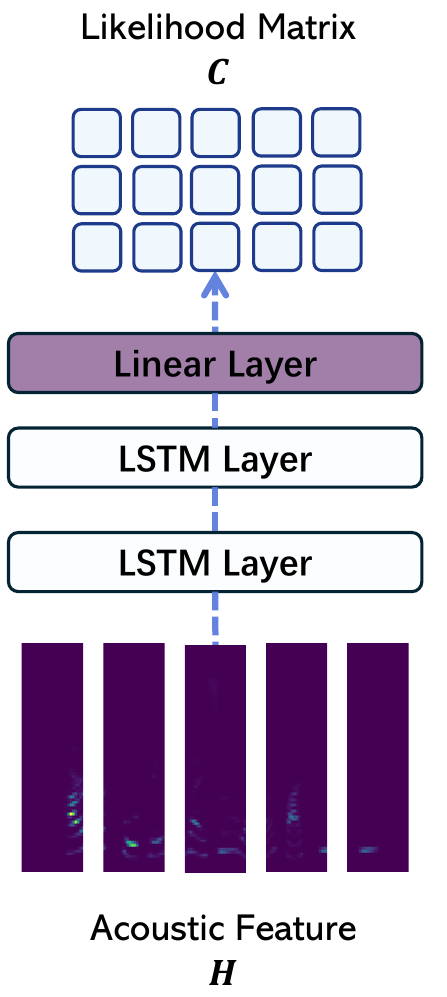}
        \caption{ASR Model}
        \label{fig:asr}
    \end{subfigure}
    \begin{subfigure}[t]{0.6\textwidth}
        \centering
        \includegraphics[width=\linewidth]{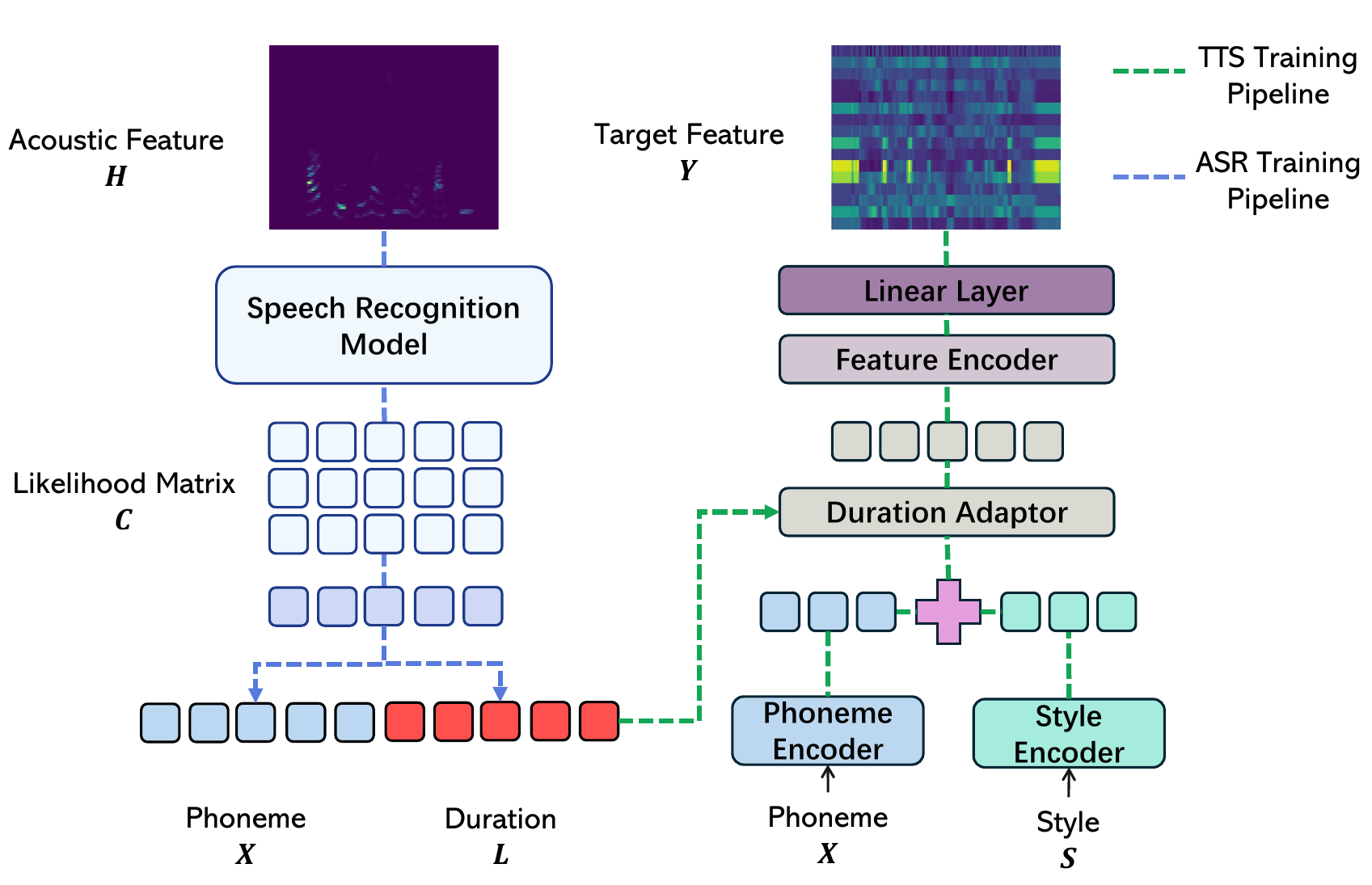}
        \caption{Aligner-Guided Training Paradigm. }
        \label{fig:agt}
    \end{subfigure}
    \caption{Architecture Diagram}
\end{figure*}

We construct a simple deep learning-based model to implement the ASR model. Figure~\ref{fig:asr} presents the architecture of the ASR model. It contains two bidirectional Long-Short Term Memory~(LSTM) layers followed by a linear layer to capture temporal dependencies in the acoustic feature and map the LSTM outputs to phoneme space to predict the likelihood of each phoneme for every frame.

The proposed ASR model cannot be directly applied to perform speech recognition tasks. The ASR model outputs a sequence of frame-level probabilities, each corresponding to a particular time step in the input signal. However, the phoneme sequence $L$ is typically much shorter, as each phoneme can span multiple frames. The mismatch between the output size $T$ and target size $L$ creates a significant challenge in aligning the predicted phoneme probabilities with the actual phoneme sequence. It leads to errors in recognizing and predicting the correct phonemes.

We impose Connectionist Temporal Classification Loss~(CTCLoss)~\cite{graves2006connectionist} to train our ASR model to address the issue of mismatch. It allows the model to learn the most likely alignment between the input frames and the target phoneme sequence without requiring explicit frame-level labels. Specifically, CTCLoss considers all possible alignments of $H$  to the target phoneme sequence and then maximizes the probability distribution of correct alignment over all possible alignments. The loss function is formally defined as: 
\begin{equation}
    \mathbf{CTCLoss} = - \mathbf{log}\sum_{\pi \in \mathbf{Align}(X,H)}\prod_{t=1}^{T}p(\pi_t|H_t)
\end{equation}
where $\pi$ represents a possible alignment between the input sequence $H$ and the target sequence $X$, $\text{Align}(X,H)$ is the set of all valid alignments, and $p(\pi_t \mid H_t)$ is the probability of the $t$-th frame in the alignment $\pi$ being assigned to the corresponding phoneme in $X$.

After training, the ASR model predicts a likelihood matrix $C \in \mathbb{R}^{(P+1) \times T}$. The additional dimension corresponds to the probability of a blank phoneme, which is typically removed in standard speech recognition tasks. However, in our method, this blank phoneme is preserved and plays a crucial role in the subsequent stages.

\subsection{Phoneme Duration Alignment}
Once the ASR model is well-trained and we have the likelihood matrix $C$. The next step is to determine the duration of each phoneme in the sequence $X$. Specifically, we need to calculate a sequence of duration $L$, where each duration $L_i$ is a non-negative integer representing the time span of the $i$-th phoneme $X_i$. This sequence must satisfy the conditions $L_i \geq 0$ and $T = \sum_{i=0}^{N-1} L_i$, where $T$ is the total number of frames in $H$.

To achieve this, we define a \textbf{P}honeme \textbf{D}uration \textbf{A}lignment (PDA) algorithm. The algorithm's objective is to collapse consecutive frames corresponding to the same phoneme and count their occurrences to determine the duration of each phoneme. The first step in this process involves applying a column-wise argmax function to the likelihood matrix $C$. This operation produces a frame-level phoneme sequence, where each entry corresponds to the phoneme with the highest probability for that frame in the acoustic feature $H$. 

Next, we iterate through the frame-level phoneme sequence. For each phoneme in the sequence, we check if it is the same as the previous phoneme or if it represents silence~$\epsilon$. If the phoneme is the same as the previous one or is a silence, we increase its duration by one. If we encounter a different phoneme, we record the duration of the previous phoneme and then reset the counter to start counting the duration of the new phoneme. This process continues until we have found the duration for each phoneme in the sequence. Figure~\ref{fig:ada} provides a clearer and more intuitive illustration of the duration alignment process.

\subsection{Aligner Guided Training}
The duration collected by the PDA algorithm are used to guide the training process of the TTS model. During training, the TTS model is provided with the phoneme sequence~$X$, the corresponding duration~$L$, and the TTS feature target~$Y \in \mathbb{R}^{M \times T}$, where $M$ represents the feature dimension and $T$ is the number of feature frames, which corresponds to the number of frames in $H$. The target feature is flexible. In this study, we follow the approach of RVAE~\cite{caillon2021rave}. We train an autoencoder to encode speech into a latent feature and use the trained decoder to reconstruct the speech. This speech-encoded latent feature serves as the TTS feature target~$Y$.

We use the StyleSpeech~\cite{lou2024stylespeechparameterefficientfinetuning} model as the backbone to build our TTS system. The StyleSpeech model provides a robust architecture that integrates subtle tonal style change into the speech generation process. It allows us to generate speech with diverse prosodic variations. In this model, the phoneme sequence~$X$ and the corresponding style sequence~$S$ are first encoded into vector embeddings. These phoneme and style embeddings are then fused to produce the acoustic embedding~$E$ by combining them through element-wise addition: $E = \mathrm{embed}(X) + \mathrm{embed}(S)$.

The sequence of acoustic embeddings $E$ is then passed through a duration adapter. The duration adapter predicts the duration~$L'$ for each acoustic embedding and duplicates each embedding~$E_i$ to match the predicted duration $L'_i$. This duration-adapted acoustic embedding is subsequently fed into a feature encoder to generate the output~$H'$. 

The TTS model is optimized using two loss functions: (1) \textbf{Speech Loss}, which minimizes the difference between the generated output and the target output using Mean-Square-Error (MSE), and (2) \textbf{Duration Loss}, which minimizes the difference between the predicted duration~$L'$ and the aligner-guided duration~$L$ using Mean-Absolute-Error (MAE).
\begin{equation}\label{eq:TTSLoss}
    \mathrm{\textbf{TTSLoss}} = \mathrm{MSE}(H,H') + \mathrm{MAE}(L,L')
\end{equation}

After applying the loss optimization, and the TTS model is trained. The generated features~$H'$ are passed through the trained decoder and produce the final generated speech~$A'$.

\begin{table*}[ht]
\centering
\begin{tabular}{lccccc}
\hline
\textbf{Model} & \textbf{WER~(\(\downarrow\))} & \textbf{WER-P~(\(\downarrow\))} & \textbf{WER-S~(\(\downarrow\))} & \textbf{MCD~(\(\downarrow\))} & \textbf{PESQ~(\(\uparrow\))} \\
\hline
StyleSpeech (Origin) & 0.361 $\pm$ 0.187 & 0.271 $\pm$ 0.180 & 0.226 $\pm$ 0.131 & 11.726 $\pm$ 3.761 & 1.057 $\pm$ 0.054 \\
\hline
With Duration Aligner \\
\hline
StyleSpeech (Latent Feature) & 0.276 $\pm$ 0.161 & 0.193 $\pm$ 0.155 & 0.162 $\pm$ 0.110 & 12.482 $\pm$ 3.995 & 1.060 $\pm$ 0.067 \\
StyleSpeech (MFCC) & 0.219 $\pm$ 0.140 & 0.130 $\pm$ 0.122 & 0.146 $\pm$ 0.108 & 11.971 $\pm$ 3.891 & 1.055 $\pm$ 0.053 \\
StyleSpeech (MelSpecs) & \textbf{0.199 $\pm$ 0.137} & \textbf{0.110 $\pm$ 0.117} & \textbf{0.131 $\pm$ 0.103} &  \textbf{11.338 $\pm$ 3.845}  & \textbf{1.068 $\pm$ 0.072} \\
\hline
\end{tabular}%

\caption{Evaluation Results of TTS systems. \textbf{(\(\downarrow\))} indicates that lower values are better, and \textbf{(\(\uparrow\))} indicates that higher values are better. The best-performing method for each metric within each training strategy is highlighted in \textbf{bold}.}

\label{tab:overall_results}
\end{table*}
\section{Experiments and Result Analysis}
\textbf{Dataset}: We use the Baker dataset~\cite{BakerDataset2020} to train and evaluate this study. We select Chinese as our evaluation language due to its complex phonetic and tonal structure. The Baker dataset contains approximately 12 hours of speech recorded using professional instruments at a frequency of 48kHz. The dataset consists of 10k speech samples from a female Mandarin speaker.

\textbf{Experimental setup}: The phoneme, style, and feature encoders each consist of four Feed-Forward Transformer~(FFT) blocks~\cite{ren2020fastspeech}. We set the dimension of the target feature to 16. For optimization, we employ a learning rate schedule with a warm-up strategy inspired by the Transformer model~\cite{vaswani2017attention}. To prevent overfitting, dropout rates are set at 0.5 for the FFT blocks and 0.1 for the length adapter.

The experiment is conducted on an NVIDIA RTX A5000 GPU using a PyTorch implementation. We select 4,000 sentences from the Baker dataset for training and 1,000 sentences for testing. The batch size is set to 64, and the model is trained for 600 epochs.

An ablation study on the training of acoustic features for the ASR model is conducted to evaluate the impact of different types of acoustic features on both the duration prediction and the overall performance of the TTS model. We trained the ASR model using MelSpec, MFCC, and latent features. The source code will be released upon acceptance of this study.

\textbf{Metrics}: We employ Word Error Rate~(WER), Mel Cepstral Distortion~(MCD)~\cite{kubichek1993mel}, and Perceptual Evaluation of Speech Quality~(PESQ)~\cite{rix2001perceptual}, to evaluate model's performance. For WER, we further evaluate the Phoneme-level WER (WER-P) and Style-level WER (WER-S). We assess the accuracy of generated speech using WER by first generating speech with the trained TTS model and then transcribing it through OpenAI's Whisper API~\cite{radford2023robust}.
\begin{figure}[t]
    \centering
    \begin{subfigure}[t]{0.14\textwidth}
        \includegraphics[width=\textwidth]{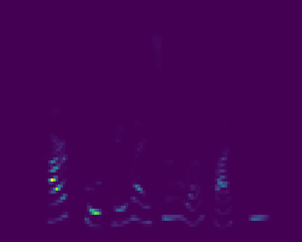}
        \caption{MelSpec}
        \label{fig:melspec}
    \end{subfigure}
    \begin{subfigure}[t]{0.14\textwidth}
        \includegraphics[width=\textwidth]{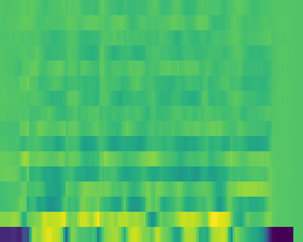}
        \caption{MFCC}
        \label{fig:mfcc}
    \end{subfigure}
    \begin{subfigure}[t]{0.14\textwidth}
        \includegraphics[width=\textwidth]{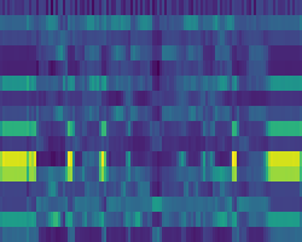}
        \caption{Latent Feature}
        \label{fig:latent}
    \end{subfigure}
    \caption{Visualization of Acoustic Features}
    \label{fig:feature_visual}
\end{figure}

\textbf{Result}: Table~\ref{tab:overall_results} presents the results of our experiment. StyleSpeech trained with aligner-guided duration shows significant improvements over models using duration labelled by external tools like the MFA, as provided in the Baker dataset. Specifically, it achieves approximately 9\% improvement when using latent features, 14\% with MFCCs, and 16\% with Mel-Spectrograms for overall WER. Additionally, for phoneme-level and tone-level WER, the model trained with aligner-guided duration shows significant improvements of around 15\% and 9\%, respectively.

In terms of the impact of different types of acoustic features, MelSpecs yield the best performance, followed by MFCCs, with latent features ranking third. All three types of acoustic features outperformed the use of duration that comes with the dataset that is labelled using external tools like MFA. 
The results indicate that the choice of acoustic features significantly affects the quality of duration prediction and, consequently, has a substantial impact on the overall performance of the TTS model. Figure~\ref{fig:feature_visual} provides a visualization of these different feature types. MelSpec exhibits a more continuous representation with clear boundaries between acoustic patterns, making it easier to detect and accurately align underlying phonemes. In contrast, MFCCs and latent features are more compact and abstract. MFCCs, being closer to the acoustic space, still maintain a reasonable level of detail for accurate alignment. Latent features, however, are more related to the latent space, which can lead to less distinct phoneme boundaries. This abstraction increases the likelihood of phoneme mismatches during alignment, resulting in weaker performance compared to MelSpec and MFCCs.

\subsection{Analysis of predicted duration}
\begin{table}[ht]
    \centering
    \begin{tabular}{lcccccccc}
        \hline
        \textbf{Phoneme} & \textbf{H} & \textbf{AO} & \textbf{H} & \textbf{AO} & \textbf{X} & \textbf{UE} & \textbf{X} & \textbf{I} \\
        \hline
        Origin   & 4  & 8  & 3  & 5  & 4  & 7  & 5  & 11 \\
        Latent   & 2  & 10 & 2  & 9  & 1  & 11 & 2  & 22 \\
        MFCC     & 5  & 6  & 5  & 6  & 2  & 8  & 3  & 20 \\
        MelSpec  & 8  & 5  & 4  & 6  & 2  & 10 & 6  & 10 \\
        \hline
    \end{tabular}
    \caption{TTS Predicted Phoneme duration}
    \label{tab:duration_results}
\end{table}
Table~\ref{tab:duration_results} presents the predicted phoneme duration for a sequence of phonemes (H, AO, H, AO, X, UE, X, I) using TTS models trained with duration generated by aligners trained using different feature types, Origin, Latent, MelSpec, and MFCC. Each row in the table represents the duration predictions from one of these features. It highlights how the choice of features impacts the alignment and timing of phonemes.

TTS model trained with Latent features shows a high variability. It tends to predict a very short duration for the initial phoneme (e.g., 'H') and an extremely long duration for the final phoneme (e.g., 'I'), as seen with values of 2 and 22, respectively. This suggests that Latent features may have a tendency to underrepresent the temporal extent of certain phonemes while overemphasizing others. It causes potentially unnatural pacing in the generated speech. The higher variability and extremities in duration predictions with Latent features may indicate difficulties in accurately capturing and aligning the phoneme boundaries.

MFCC feature shows moderate variability, with predicted duration ranging from 2 to 20. While MFCCs provide reasonable accuracy in phoneme duration prediction, the presence of outliers, such as the duration of 20 for the final phoneme, suggests occasional misalignment, which might affect the naturalness of the generated speech. MelSpec feature results in more balanced duration predictions across the phonemes, ranging from 2 to 10. This distribution suggests that MelSpec provides a better representation of phoneme length. It facilitates more natural and consistent speech output. The consistent duration predictions across different phonemes indicate effective alignment, which likely contributes to MelSpec's superior performance in TTS models.

\section{Conclusion}
In this study, we validate the significance of duration in TTS training and propose an effective aligner-guided training paradigm. Our experiments show that Mel-Spectrograms outperform other acoustic features. These contributions provide valuable insights into optimizing TTS systems for more natural and accurate speech generation.

\newpage

\bibliographystyle{IEEEbib}
\bibliography{reference}

\vspace{12pt}

\end{document}